\documentstyle[12pt]{article}

\newcommand{\be}{\begin{equation}}
\newcommand{\ee}{\end{equation}}
\newcommand{\beqn}{\begin{eqnarray}}
\newcommand{\eeqn}{\end{eqnarray}}

\begin{document}
\rightline{DESY 95-124}
\rightline{hep-ph/9507271}
\bigskip
\begin{center}
\begin{large}
{\bf Nonsinglet Contributions to the Structure Function $g_1$
at Small-x} \\
\vspace{1cm}
J. Bartels\\
\end{large}
{\it II. Institut f\" ur Theoretische Physik, Universit\" at Hamburg}
\\
\vspace{1cm}
\begin{large}
B.I.Ermolaev\footnote{The research described in this paper has been
made possible in part by the Grant R2G 300 from the International
Science Foundation and the Russian Government. This work has also been
supported by the Volkswagen-Stiftung.}
\\
\end{large}
{\it A.F.Ioffe Physical-Technical Institute, St.Petersburg,
194021, Russia}
\\
\vspace{1cm}
\begin{large}
 M. G. Ryskin\footnote{Work supported by the Grant INTAS-93-0079
and by the Volkswagen-Stiftung}
\end{large}
\\
{\it St.Petersburg Nuclear Physics Institute, Gatchina, St.Petersburg,
188350, Russia}
\\
\end{center}
\vspace{1cm}

{\bf Abstract:}
\noindent
Nonsinglet contributions to the $g_1(x,Q^2)$ structure function are
calculated in the double-logarithmic approximation of perturbative
QCD in the region $x \ll 1$. Double logarithmic contributions
of the type $(\alpha_s \ln ^2 (1/x))^k$ which are not included in
the GLAP evolution equations are shown to give a stronger rise at
small-x than the extrapolation of the GLAP expressions. Further
enhancement in the small-x region is due to non-ladder Feynman graphs
which in the DLA of the unpolarized structure functions do not
contribute. Compared to the conventional GLAP method (where neither
the whole kinematical region which gives the double logs nor the
non-ladder graphs are taken into account) our results lead to a growth
at small-x which, for HERA parameters, can be larger by up to factor of
10 or more.

\newpage
\section{Introduction}
\setcounter{equation}{0}
In the framework of perturbative QCD the theoretical investigation of
deep inelastic scattering in the HERA regime puts particular emphasis
on the calculation of structure functions in the region $x \ll 1$.
So far, main attention has been given to the small-x behavior of the
gluon structure function: from the theoretical
point of view, the strong rise predicted by perturbative QCD violates
unitarity and hence requires corrections which restore unitarity.
Experimentally, such a rise has been observed, and it is attractive
to interpret it as a manifestation of the BFKL Pomeron ~\cite{BFKL}.\\
\\
 From the theoretical side, however, also the fermion structure
functions have quite an interesting small-x behaviour. Recently
{}~\cite{EMR} it has been shown, for the case of the flavor nonsinglet
contribution to $F_1$, that the small-x behavior of quark
structure functions is stronger than
what one would obtain from simply extrapolating the GLAP ~\cite{GL,AP}
evolution
equations into the small-x region. The reason for this lies in the fact
that at small-x a new region in phase space opens up which gives rise
to logarithms of the type $(\alpha_s \ln^2 (1/x) )^n $ and which
is not taken into account within the GLAP method. As long as we are
are dealing only with neutral currents and unpolarized structure
functions,
this observation may seem to be slightly academic, since the gluons in
the flavor singlet will dominate at small x, and the quarks represent a
nonleading effect. The situation, however, changes drastically if one
considers, for example, polarized structure functions and their sum
rules ~\cite{review}. It is well known ~\cite{AP} that the polarized
gluon structure function is no longer growing as $(1/x)^{\lambda}$
with $\lambda$ close to 1, but rather with a $\lambda$ close to 0.
Consequently, the polarized gluons are no longer dominating over the
fermions, and the small-x behavior of the quarks becomes as vital for
polarized deep inelastic structure functions as that of the gluons.\\ \\
It may be useful to briefly review the small-x behavior of the
(unpolarized) gluon
structure function in the GLAP scheme and in the BFKL Pomeron. As it
is well known, GLAP sums logarithms of the type
$( \alpha_s \ln (Q^2/\mu^2) )^k$,
whereas BFKL keeps terms of the form
$( \alpha_s \ln (1/x) )^k$. In the region of mutual overlap where the
DLA is valid one is  summing terms of the form
$( \alpha_s \ln (Q^2/\mu^2) \ln (1/x) )^k$. In terms of the anomalous
dimension $\gamma_{gg}$ of the two-gluon operator, the GLAP formalism
uses a fixed order (one or two-loop) in $\alpha_s$ expression. As
a function of $n$, $\gamma_{gg} (n)$ is singular near $n=1$
($n$ is the moment index which in the small-x region coincides with
angular momentum $j$ of the cross channel). In contrast,
the BFKL Pomeron provides all singular terms of the form $\sum
(\alpha_s/(n-1))^k c_k$. Finally, in a physical
gauge both GLAP and the
BFKL Pomeron are sums of ladder diagrams, but the phase space
for the transverse momenta of the gluons has, in the GLAP case, the
property of strong ordering, whereas in the BFKL approximation this
feature is almost completely lost \footnote{During the rapidity
evolution of the BFKL Pomeron, the mean value of the logarithm of the
transverse momentum is still growing, but the fluctuations are much
larger than in the GLAP case.}.\\ \\
This situation changes if we consider fermions at small x ~\cite{EMR}.
It has been observed almost 30 years ago ~\cite{GGFL}
that in the Regge limit scattering amplitudes with two fermions in the
t-channel have two powers of $\ln(s/\mu^2)$ per loop, i.e. one is
summing
terms of the form $(\alpha \ln^2 (s/\mu^2) )^n$ rather than single
logs. These double-logs cannot be reached within the GLAP scheme which,
at small x, only includes the (less important) terms of the form
$( \alpha_s \ln (Q^2/\mu^2) \ln (1/x) )^n$. This difference is
seen very clearly if one analyses the relevant region of integration
inside the fermion ladders: GLAP has the same ordering of transverse
momenta as
in the gluon ladders, but the double-logs ~\cite{GGFL} come from a
region with no ordering in $k_T$ or the angle $\theta$. Only the
longitudinal components of the Sudakov variables are still ordered.
In other words, in addition to the GLAP region there is another part
of the phase space which contributes to these double-logs.
Finally, it is instructive to describe this difference also
in terms of the anomalous dimension. Within the GLAP analysis the
anomalous dimension of the quarks, as a function of the moment
index $n \approx j$, has a pole near n=0. In the
Regge analysis which includes the double-logs, one obtains a
quite different dependence upon n; only for
$\sqrt{\alpha_s} \ll n \ll 1$ one recovers the usual result. The small-x
behavior, on the other hand probes the region $n \sim \sqrt{\alpha_s}$,
where the two methods lead to different predictions. \\ \\
In this paper we will calculate the small-x behavior of the nonsinglet
contribution to the polarized structure function $g_1$; in a
forthcoming paper we will extend our analysis to the flavor singlet
part which contains mixing between the quarks and the gluons.
We find that the small-x behavior is stronger than the extrapolation
of the GLAP formula predicts. In comparison with the fermions in the
unpolarized structure function ~\cite{EMR} the polarized quark
structure functions have another new property. In contrast to $F_1$
where a simple ladder structure gave the relevant logarithms, the
polarized structure function $g_1$ receives contributions also from
non-ladder graphs. Our analysis will show that these terms lead to a
further enhancement at small-x.\\ \\
This paper is organized as follows. In section 2 we briefly review the
definitions of the structure functions, and we discuss their signature
properties by studying the double logarithmic contributions of the
first two loop corrections.
In section 3 we construct the infrared evolution equation for $g_1$
and find its solution. The final section 4 contains a brief discussion
of our results.

\section{Signatures of the Structure Functions}
\setcounter{equation}{0}

The standard parametrization of the hadron tensor $W_{\mu \nu}$ of
deep inelastic scattering (neglecting the contribution of weak currents)
has the form ~\cite{IKL}:
\beqn
W_{\mu \nu}
  = (-g_{\mu \nu} + \frac{q_{\mu} q_{\nu} } {q^2})F_1
  + (p_{\mu} - q_{\mu} \frac{pq}{q^2} )
    (p_{\nu} - q_{\nu} \frac{pq}{q^2} ) \frac{F_2}{pq}
         \nonumber \\
      + \imath \epsilon_{\mu \nu \alpha \beta} q^{\alpha} s^{\beta}
           \frac{m}{pq} g_1
  + \imath \epsilon_{\mu \nu \alpha \beta} q^{\alpha}
             (s^{\beta} - p^{\beta} \frac{sq}{pq} ) \frac{m}{pq} g_2
\eeqn
where $p$, $s$, $m$ are the four momenta of the incoming parton (quark),
its polarization vector and its mass, resp., and $q$ the lepton
momentum transfer ($-q^2=Q^2$). The structure functions $F_1$, $F_2$,
$g_1$, and $g_2$ depend upon $Q^2$ and $x=Q^2/2pq$.\\ \\
The structure functions in (1) are energy discontinuities of scattering
amplitudes for the elastic Compton scattering of a virtual photon off
the proton. In analogy with (1) we write:
\beqn
T_{\mu \nu}& = & \imath \int d^4x e^{\imath
qx}       <N|T \left(J_{\mu} (x) J_{\nu}(0) \right)| N>
   \:\:\:\:\:\:\:\:\:\:\:\:\:\:\:\:\:\:\:\:\:\:\:\:\:\:\:\:\:\:\:\:\:
       \nonumber \\
           & = &
    (-g_{\mu \nu} + \frac{q_{\mu} q_{\nu} } {q^2})T_1
  + (p_{\mu} - q_{\mu} \frac{pq}{q^2} )
    (p_{\nu} - q_{\nu} \frac{pq}{q^2} ) T_2
         \nonumber \\
&   &   + \imath \epsilon_{\mu \nu \alpha \beta} q^{\alpha} s^{\beta}
           \frac{m}{pq} T_3
  + \imath \epsilon_{\mu \nu \alpha \beta} q^{\alpha}
             (s^{\beta}(pq) - p^{\beta} (sq) ) \frac{m}{pq} T_4
\eeqn
with
\beqn
F_1= -\frac{1}{2 \pi} Im T_1 \,\,\,\;\;\;\; F_2=-\frac{pq}{2\pi}ImT_2
\eeqn
\beqn
g_1 = -\frac{1}{2 \pi} Im T_3 \,\,\,\;\;\;\; g_2=-\frac{pq}{2\pi}ImT_4
\eeqn
It is easy to see that these amplitudes satisfy the following crossing
symmetry relations: $T_1,T_2$ and $T_4$ are symmetric with respect to
the replacement $s\to -s$ (i.e. $s\to u$ channel), while the amplitude
$T_3$ changes sign. Indeed, the tensor $T_{\mu \nu}$ is symmetric
under the interchange of $\mu$ and $\nu$ and $q \to -q$. But $q \to -q$
means $x \to - x$, i.e. $s \to u$. Since the tensors in front of
$T_1$ and $T_2$ are symmetric, $T_1$ and $T_2$ must be even under
$s \to u$. Correspondingly, $T_3$ must be odd, and $T_4$ whose tensor
has an additional power of $q$, is even again.
In other words, $T_1,T_2,T_4$ are the amplitudes of the positive
signature, and $T_3$ has the negative signature.
\footnote{All the other kinematical factors (like $pq$) which sometimes
enter into the definition of the structure functions do not change the
signature (i.e. crossing symmetry) properties of a given amplitude.}
In the following we shall concentrate on this odd signature case $T_3$
which gives $g_1$.
\\ \\
One can check the main implications of the signature assignment
by simply calculating the amplitudes in lowest order  perturbation
theory. In the Born approximation they take the form:
\beqn
T^{(0)}_1=e^2_q\left (\frac s{s-Q^2+\imath\epsilon}+\frac {-s}{-s-Q^2+\imath
\epsilon}\right) ,
\\
T^{(0)}_2=e^2_q\frac{4Q^2}{s^2}\left (\frac s{s-Q^2+\imath\epsilon}+
\frac {-s}{-s-Q^2+\imath\epsilon}\right) ,
\\
T^{(0)}_3=e^2_q\left (\frac s{s-Q^2+\imath\epsilon}-\frac {-s}{-s-Q^2+\imath
\epsilon}\right) ,
\eeqn
\be
T^{(0)}_4=0
\ee
($e_q$ is the electric charge of the initial quark of flavour $q$.)
We have deliberately chosen this particular way of writing eqs.(5-8) in
order to stress that $T_1$ and $T_2$ are symmetric under the
replacement $s\to -s$, and $T_3$  is antisymmetric. \\ \\
Now let us consider the radiative corrections to the $T_i$.
In the one-loop approximation there are two Feynman graphs yielding
DL-contributions to $T_i$  (see Fig.2). However, since ultimately we
are interested in the discontinuity of the amplitude, we have to
consider, besides the DL-terms, also the $\imath\pi$-parts of the
energy logarithms. When $s>0$, only the graph 2a yields the
$\imath\pi$-contribution. The result is:
\be
T^{(1)}_1=e^2_q\frac{g^2C_F}{16\pi^2}\left[ ln^2(\frac{-s}{\mu^2})+ln^2\frac{s}
 {\mu^2}-2ln^2\frac{Q^2}{\mu^2}\right],
\ee
\be
T^{(1)}_2=2xT^{(1)}_1,
\ee
\be
T^{(1)}_3=e^2_q\frac{g^2C_F}{16\pi^2}\left[ ln^2(\frac{-s}{\mu^2})-ln^2\frac{s}
{\mu^2}\right],
\ee
\be
T^{(1)}_4=0
\ee
where $C_F=(N^2-1)/2N$ for the colour group SU(N), $g$ is the
QCD-coupling constant which we will keep fixed within the DLA,
and $\mu$ is an infrared mass scale for the transverse momentum:
\beqn
            \mu < k_t
\eeqn
Eqs.(9-11) show that the one-loop expressions, $T^{(1)}_i$, for the
functions $T_i$ still have the same signatures as the Born terms.
As a result, in $T_1$ and $T_2$ the double logs add up, whereas
in $T_3$ only a term $\sim \imath \pi ln(s/\mu^2)$ survives.
For the energy discontinuity, in both $T_1$ and $T_3$ the
$\imath \pi$ pieces in $\ln ^2 (s/\mu^2)$ give the relevant
contribution. This pattern will not change if we include more rungs
in higher orders in $g^2$, i.e. when building the usual sum of ladder
diagrams. \\ \\
However, non-ladder Feynman diagrams can also contribute to $T_i$ in
the DLA, and their role is quite different for the even- and
odd-signature amplitudes. For example, from the case of the elastic
scattering of quarks it is known ~\cite{KL} that non-ladder graphs
contribute differently to positive and negative signatured amplitudes.
In our case, the non-ladder diagrams first appear (within the DLA)
in the two loop approximation, i.e. in the order $g^4$ of
perturbation theory. They are illustrated in Fig.3.
Let us study these contributions in more detail. In particular
we want to demonstrate and explain that they do contribute to $T_3$ but
not to $T_1$ or $T_2$. \\ \\
Following the line of arguments introduced
in ~\cite{GGFL}, each of the non-ladder diagrams can be viewed as being
obtained from a ladder graph by adding non-ladder virtual gluons to it.
In order to give the double logarithm, these gluons couple only to the
side lines of the ladder (i.e. not to the rungs), and they must be
softer (i.e.have smaller $k_t$) than all those ladder gluon rungs
which are embraced by the non-ladder gluon. Otherwise they would destroy
double logs of the ladder. Morover, those propagators of the ladder
to which the non-ladder gluon is attached, must be close to the mass
shell - its virtuality should be smaller than the transvers momentum
square of the non-ladder gluon. In the following, these
non-ladder type gluons will be called bremsstrahlung gluons.
The important consequence of their
``softness`` is that, in the numerator of the expession for the
non-ladder graph, the momenta of the bremsstrahlung gluons can
be neglected compared to the ladder gluon momenta. Furthermore, the
momenta along the side lines of the ladder, are small in comparison
with the initial momenta $q$ and $p$. In Fig.3 the momentum of the
bremsstrahlung gluon is denoted by $k_2$.
The remaining momenta in all these numerators give, for each
diagram, the factor $\pm 4k^2_1\cdot pk_1$ (where the sign plus stands
for the diagrams in Fig.3a,b, while the minus sign belongs to
Figs.3c,d). In front of this numerator we have a tensor which is
built from $p_{\mu}$, $q_{\mu}$ or $g_{\mu \nu}$.
These tensors are identical for the graphs (a,b,c,d), and those
of the (a',b',c',d')-diagrams differ by the replacement
$\mu\rightleftharpoons\nu$ and $q\to -q$.\\  \\
The way in which these double logs sum up or cancel, depends upon
color and signature. First it is easy to see that for a color singlet
t-channel different bremstrahlung gluons cancel: for example,
in Fig.3a and c the bremsstrahlung gluons are emitted from the lower
left external line, and we are summing over different end points.
Since we have color zero in the t-channel, both contributions cancel.
As a result, in this order all double log bremsstrahlung
contributions in Fig.3 cancel
(this will change when we go to higher order, as we shall discuss
further below).
\\ \\
Signature becomes crucial when we consider the $\imath \pi$ terms
which are essential for the s-discontinuity. To be more precise,
starting from the order $g^4$ where the nonladder graphs appear,
even and odd signature amplitudes start to behave quite differently.
For example, the graphs in Figs.3a' and
b' have no s-cuts, but those in c' and d' can be redrawn as shown in
Fig.4, and therefore they contribute to the s-discontinuity. Combining
Figs.3a,b with c',d' one finds cancellation for even signature,
non-cancellation for odd signature. As a result, the non-ladder
graphs contribute to $T_3$, but not $T_1$ or $T_2$. This is in
accordance with the conclusion of ~\cite{KL} for the quark scattering
amplitude. \\ \\
When we move on to higher orders in $g^2$, most of these features
can easily be generalized. For the remainder of this section we
remove the photon lines at the upper end, and we restrict ourselves
to quark scattering. First let us return to the bremstrahlungs
gluons. Starting from a fermion ladder with a few gluon rungs, each
non-ladder gluon will comprise a subset of the rungs. In order to
give the maximal number of double logarithms, this bremsstahlung
gluon must be softer (i.e.have smaller $k_t$) than any line of
this subset of rungs; at the same time, the fermion lines to which
the bremsstrahlung gluon couples must be softer than the gluon itself.
This implies that the bremsstrahlung gluons form a ``nested`` structure.
The summation of all these graphs is done as described in ~\cite{KL}.
Consider a typical diagram; pick the internal line with the lowest
$k_t$. If it is a fermion line, the amplitude can be decomposed as
indicated in Fig.5a. If, on the other hand, it is a gluon line, it
must be a bremstrahlungs gluon which - according to Gribov's theorem
 - ~\cite{Gri} must go from one external leg to another (Fig.5b).
The sum of all diagrams, therefore, can be written as shown in
Fig.5c; the first term on the rhs denotes  the Born term.
If we take the derivative with respect to the infrared
cutoff $\mu^2$ (which appears only in the line with the smallest
momentum), then we arive at the infrared evolution equation
illustrated in Fig.5c (whether or not the Born term contributes to this
derivative has to be decided from case to case). This is the
infrared evolution equation which has proven to be an efficient
instrument for the investigation of the elastic scattering of quarks
{}~\cite{KL}, and also in inelastic reactions at high energies
{}~\cite{EL}.\\ \\
The presence of the second term on the rhs has an important consequence.
Even we restrict ourselves to amplitudes with color zero in the
t-channel, the colored gluon forces the four quark amplitude to be
in a color octet. Therefore, this second part appears as an
inhomogeneous
term in the nonlinear evolution equation, and we have to solve a
separate evolution equation for the color octet amplitude.
Furthermore, in the case of a colour singlet amplitude this term contributes
only to the odd signature,
as it can be seen as follows. Consider, for example,
the two diagrams shown in Fig.6a and b. For the odd-signature amplitude,
they both come with the opposite sign. Redraw the first diagram
as illustrated in Fig.6c, and note that the upper line to which
the soft gluon couples has the opposite direction: this gives an
additional minus sign from the color factor, and the two diagrams
sum up. Conversely, in the even signature case they cancel.
As a result, the nonplanar diagrams appear only in the odd signature
amplitude. \\ \\
Finally, the infrared evolution equation for the (even signature)
octet amplitude. Following the same arguments as before, we are led
to the nonlinear evolution equation whose structure we illustrate in
Fig.7. In the second term, only the even signature color octet
contributes: the gluon has negative signature, and for the subamplitude
the even signature configuration gives the maximal number of double
logs. Combining this with the color factors, one finds that only the
color octet survives.

\section{Infrared Evolution Equation for $g_1$}
\setcounter{equation}{0}

In this section we apply the methods outlined in the previous
section in order to construct and solve the infrared evolution equation
for the nonsinglet contribution to $g_1$. We will take into account
both the DL-contributions and the $i \pi$ contributions. \\ \\
Let us consider, again, the forward elastic photon-quark scattering
amplitude $T_{\mu \nu}$ in DLA. As before, $q$ and $p$ are the
(collinear) four
momenta of the photon and the quark, resp., and we are in the
regime where $s \gg -q^2=Q^2 \gg p^2$. The structure of the infrared
evolution equation for quark scattering has been discussed in the
previous section, and we now apply the same ideas to photon quark
scattering. The equation is illustrated in Fig.8,
and its origin is easily understood. The upper blob on the rhs denotes
the photon-quark scattering amplitude with the lower (incoming) legs
having transverse momentum square equal to the infrared cutoff $\mu^2$.
The lower blob on the rhs represents a quark quark elastic scattering
amplitude with $k_T^2 = \mu^2$ for all external momenta.
In order to translate this into an
explicit equation, we note that $T_{\mu \nu}$ depends upon
the two variables $z=\ln(s/\mu ^2)$ and $y=\ln (Q^2/\mu^2)$. Therefore,
the lhs of Fig.5 has the two terms:
\beqn
- \mu^2 \frac{\partial T_{\mu \nu} }{ \partial \mu^2}
    = \frac{\partial T_{\mu \nu}}{\partial z}
    + \frac{\partial T_{\mu \nu}}{\partial y}  .
\eeqn
The convolution on the rhs becomes simpler if we write our amplitudes
in the Mellin representation w.r.t. $s$ ($z=\ln (s/\mu^2)$):
\beqn
T_i(z,y) = \int_{-i\infty}^{i\infty} \frac{d\omega}{2 \pi i}
(\frac{s}{\mu^2} )^{\omega} \xi^{(i)} R_i(\omega, y),
\eeqn
where $\omega$ denotes the angular momentum $j$, and $\xi$ is the
signature factor:
\beqn
\xi^{(i)} = \frac{ e^{-i \pi \omega} + \tau_i}{2}
\eeqn
with $\tau_3 = -1$. As usual, the integration contour runs to the
right of all singularities in the $\omega$-plane.
A similar representation is used
for the quark-quark scattering amplitude, and the transform is denoted
by $f_0^{(-)}$. The infrared evolution equation of Fig.8 then takes the
form:
\beqn
\omega R_i + \frac{\partial R_i}{\partial y}
                   = \frac{1}{8 \pi^2} R_i f_{0}^{(+)}
\eeqn
for i=1,2 and
\beqn
\omega R_3 + \frac{\partial R_3}{\partial y}
                   = \frac{1}{8 \pi^2} R_3 f_{0}^{(-)}
\eeqn
for the amplitude $T_3$. Eq.(3.4) has been discussed and solved in
{}~\cite{EMR}. \\ \\
In order to solve eq.(3.5) for $R_3$, we need to know $f_{0}^{(-)}$.
This is the quark scattering amplitude discussed in the previous
section. It satisfies the evolution equation of Fig.5 ~\cite{KL}:
\beqn
f_0^{(-)} (\omega) = \frac{N^2-1}{2N} \frac{g^2}{\omega}
      -\frac{N^2-1}{N} \frac{g^2}{4 \pi^2}
             \frac{1}{\omega^2} f_8^{(+)} (\omega)
      + \frac{1}{8 \pi^2 \omega} \left( f_0^{(-)} (\omega) \right)^2
\eeqn
The second term on the rhs is due to the signature changing
contributions of gluon bremsstrahlung which lead us to define
an even-signature quark quark scattering amplitude $f_8^{(+)}(\omega)$
with color octet quantum number in the t-channel. This amplitude has
the infrared evolution equation shown in Fig.7:
\beqn
f_8^{(+)} (\omega) = -\frac{g^2} { 2N \omega}
     +  \frac{N g^2}{8 \pi^2 \omega} \frac{d}{d\omega} f_8^{(+)}
             (\omega)
     +\frac{1} {8 \pi^2 \omega} \left( f_8^{(+)} (\omega) \right)^2
\eeqn
Its solution follows from the discussion given in ~\cite{KL}: using the
transformation
\beqn
f_8^{(-)} (\omega) = Ng^2 \frac{\partial}{\partial \omega}
           \ln u(\omega)
\eeqn
the Riccati equation (3.7) is equivalent to the linear differential
equation
\beqn
\frac{du^2}{dz^2} - z \frac{du}{dz} - \frac{1}{2N^2} u = 0
\eeqn
where
\beqn
z=\omega/\omega_0, \,\,\, \omega_0 = \sqrt{N g^2/8 \pi^2}.
\eeqn
After
a trivial transformation this equation is solved by a parabolic
cylinder function. As a result, $f_8^{(+)}$ has the form:
\beqn
f_8^{(+)}(\omega)
         & = & Ng^2 \frac{d}{d\omega} \ln \left( e^{z^2/4} D_p (z)
                              \right)
       \nonumber \\
        & = & Ng^2 \frac{d}{d\omega} \ln
    \left(   \int_0^{\infty} dt t^{-1-p} e^{-zt-t^2/2}  \right)
\eeqn
with
\beqn
p=-1/2N^2.
\eeqn
With this we return to (3.6) and obtain for $f_0^{(-)}$:
\beqn
f_0^{(-)} = 4 \pi^2 \omega \left( 1 - \sqrt{ 1 - \frac{g^2(N^2 - 1)}
            {4 \pi^2 N \omega^2} [1 - \frac{1}{2 \pi^2 \omega}
                                     f_8^{(+)} (\omega)]
                                                       } \right)
\eeqn
(the minus sign in front of the square
root follows from the requirement that, for large $\omega$, the solution
has to match the Born approximation). \\ \\
Having found the function $f_0^{(-)} (\omega)$, we are now able to solve
eq.(3.5):
\beqn
R_3(\omega, y) = C(\omega)
            e^{( -\omega + f_0^{(-)} (\omega) /8\pi^2 ) y}
\eeqn
with some unknown function $C(\omega)$. This function is determined if
we impose the matching condition ~\cite{EMR}:
\beqn
T_3 (z, y=0) = \tilde{T_3} (z)
\eeqn
where $\tilde{T_3}(z)$ is the corresponding invariant amplitude with a
(nearly) on-shell photon: $-q^2=\mu^2$. Since this amplitude no longer
depends upon $Q^2$, its infrared evolution equation (which is the
analogue of eq.(3.5))) has no $y$-derivative. On the other hand, its
Born term now has a dependence upon $\mu^2$ which is no longer
negligeable and leads to a nonzero contribution on the rhs (Fig.8).
Denoting the Mellin transform of $\tilde{T_3}$ by $\tilde{R_3}$
the matching condition (3.15) simply becomes $C(\omega) =\tilde{R_3}$,
and the equation for $\tilde{R_3}$ reads:
\beqn
\omega \tilde{R_3} = c_3 + \frac{1}{8 \pi^2} \tilde{R_3} f_0^{(-)}.
\eeqn
The boundary condition for $\tilde{T_3}$ is:
\beqn
\tilde{T_3} (z=0) = \tilde{T}_{3 \, Born} = c_3 \nonumber
\eeqn
\beqn
\tilde{R_3} = \frac{c_3}{\omega}
\eeqn
where $\tilde{T}_{3 \, Born}$ is given in (2.7) and leads to
$c_3=2 e_q^2$.\\ \\
Inserting the solution for $\tilde{R_3}=C(\omega)$ into (3.15) we
obtain:
\beqn
T_3 (z, y)
    = 2 e_q^2 \int_{i \infty}^{i \infty}  \frac{d \omega}{2 \pi i}
          \left( \frac{s}{Q^2} \right)^{\omega} \xi^{(-)}
          \frac{1} {\omega - f_0^{(-)} (\omega) / 8\pi^2}
          e^{ y f_0^{(-)} (\omega) /8 \pi^2}
\eeqn
In order to obtain our final result, $g_1$, we have to take the
discontinuity in $s$. With the signature factor $\xi^{(-)} = \imath
\pi \omega /2$ and the variables $x=Q^2/s$ and $Q^2$
we arrive at
\beqn
g_1(x, Q^2) = \frac{e_q^2}{2} \int_{-i\infty}^{i \infty}
                \frac{d \omega}{2 \pi i}
 x^{-\omega} \left( \frac{Q^2}{\mu^2} \right)^{f_0^{(-)}(\omega)/8\pi^2}
                \frac{\omega}{\omega - f_0^{(-)}(\omega) /8 \pi^2}
\eeqn
In the limit where $1/x$ is much larger than $Q^2/\mu^2$, the
leading contribution comes from the rightmost singularity in the
$\omega$-plane.

\section{Discussion}
\setcounter{equation}{0}

The most interesting aspect of our result (3.19) is the small-x
behaviour. The leading singularity in the $\omega$-plane is the branch
point due to the vanishing of the square root in $f_0^{(-)}$ in
(3.13). As a first approximation, let us take the number of colours
 $N$ to be large.
As it has been discussed in ~\cite{KL}, $f_8^{(+)}$ can then be
approximated by the first term on the rhs of (3.7) (i.e.its Born
term), and the location of the branch point follows from the condition:
\beqn
 0=1 - \frac{g^2 N}
            {4 \pi^2  \omega^2} [1 + \frac{g^2} {4 \pi^2 N \omega^2}]
\eeqn
(note that our $\omega = j$, i.e. it differs from the $\omega= j-1$
which is commonly used in the context of the BFKL Pomeron and the
unpolarized gluon structure function).
Expanding in inverse powers of $N$ we find for the first two terms:
\beqn
\omega = \omega^{(-)} = \omega^{(+)} (1 + \frac{1}{2N^2})
\eeqn
where
\beqn
\omega^{(+)} = \sqrt{\frac{g^2 (N^2-1)}{4N \pi^2}}
\eeqn
is the rightmost singularity for the flavor nonsinglet part of $F_1$
found in ~\cite{EMR}. Lut us stress
 that the leading singularity of the negative
signature amplitude ($g_1$) is shifted to the right  compared to
the the positive signature one $\omega^{(+)}$~\cite{KL}.
 Fistly such a behaviour of the negative signature amplitudes were noticed
in~\cite{GLN} for the elastic scattering in QED.\\
A more accurate determination of the branch point
singularity leads, instead of (4.2), to:
\beqn
\omega^{(-)} & \approx & \omega^{(+)} \cdot 1.04
       \nonumber \\
       & \approx & 0.41
\eeqn
for $\alpha_s = 0.18$. As a result, the power of $1/x$ of the
nonsinglet piece of $g_1$ is stronger than that of $F_1$ by about 4 \%.
This enhancement has its origin in the non-ladder graphs which where
absent in the calculation of $F_1$. In the HERA region, the enhancement
of $g_1$ relative to $F_1$ ~\cite{EMR} is approximately
\beqn
g_1/F_1 & \approx & 1.+ \omega^{(+)} 0.04 [\ln (1/x) + \frac{1}{2} \ln
                        (Q^2/\mu^2) ]
\nonumber \\
         & \approx & 1.13
\eeqn
for $x=10^{-3}$, $Q^2 = 20 GeV^2$, and $\mu^2=4 GeV^2$.
This estimate may serve as a crude method of estimating the non-singlet
$g^{n.s.}_1$ in the
small-x region, once $F^{n.s.}_1$ has been measured.
\\ \\
Let us finally see how our result (3.19) is related to the
GLAP expression. To this end we take $Q^2/\mu^2$ to be much larger
than $1/x$; the exponent $f_0^{(-)}(\omega)/8 \pi^2$ plays the role of
the anomalous dimension
$\gamma_{qq}$. For $\sqrt{\alpha_s} \ll \omega \ll 1$ (i.e.
$g^2/\omega^2$ is small) we obtain from (3.13):
\beqn
\gamma_{qq} (\omega) = 2 \frac{\alpha_s}{4\pi \omega} \frac{N^2-1}{2N}
\eeqn
which agrees with the singular part of the quark anomalous dimension.
In a straightforward extrapolation of the GLAP approximation
into the small-x region ~\cite{ES,BFR,CR} one would use this
singular term in the anomalous dimension
and performe a saddle point analysis. This has lead to the
conclusion that at small x $g_1 \sim \exp( \sqrt{const \ln (Q^2/
\Lambda^2) \ln(1/x)} )$, in analogy with the small-x behavior of the
gluon structure function (at fixed $\alpha_s$). In contrast to this,
our result in (3.19)
gives a quite different answer for the $\omega$ dependence of the quark
anomalous dimension and hence for the small-x behavior of $g_1$.
For a numerical estimate of the difference we simply combine (4.5) with
the result for $F_1$ which was obtained in ~\cite{EMR} (note that for
the flavor nonsinglet case the $\gamma_{qq}$ is the
same for the polarized and the unpolarized structure function):
for the non-singlet quark structure function with the initial condition
$q(x, Q_0^2) \propto \delta(1-x)$
\footnote{It does not seem very plausible to assume that $g_1(x, Q_0^2)$
is singular at $x \to 0$, therefore we think that the somewhat
oversimplified $\delta$-function ansatz is justified.}
a comparison has been made of the GLAP-evolution
at small-x and the double logarithmic formula derived in ~\cite{EMR},
and a difference
of up to a factor of 10 (at typical HERA values for $x$ and $Q^2$)
was found. Together with (4.5), this leads to a slightly stronger
discrepancy in $g_1$: a factor of ten or even more. \\ \\
An even stronger discrepancy exists between our result and the
Regge prediction ~\cite{H,EK,BL}. For the flavor nonsinglet (isotriplet)
exchange the small-x behavior is given by $a_1$ exchange. It
has the form $(1/x)^{\alpha_R}$, and the $a_1$ intercept $\alpha_R$ is
believed to lie somewhere between $-0.5$ and $0$ ~\cite{EK}, i.e. the
flavor nonsinglet part of $g_1$ is predicted to be regular as $x \to
0$!  \\ \\
Having found that the double logarithms in the small-x region
are numerically important, we have to adress the slightly more
involved analysis of the flavor singlet case. Here we are facing the
mixing between gluons and quarks: as it can be seen from ~\cite{AP}
or, more directly, from ~\cite{ES,BL}, the small-x behavior of the
polarized gluon structure function is less singular than the
unpolarized one and, therefore, competes with the quarks. This analysis
is in progress, and results will be presented in a forthcoming paper.\\
\\
{\bf Acknowledgements:} We are grateful to E.A.Kuraev, M.M Terekhov,
and, especially, to L.N.Lipatov for useful discussions. Two of us
(B.E. and M.R.) wish thank DESY for their hospitality.
\newpage
\noindent
{\bf Figure Captions:} \\ \\
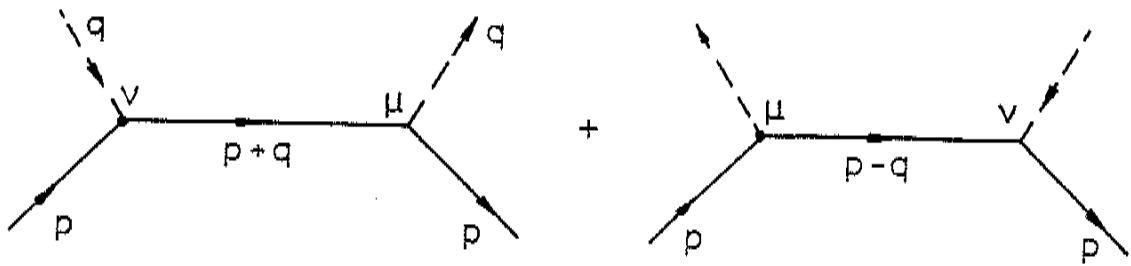
Fig.1: The Born approximation for $T_{\mu \nu}$.\\ \\
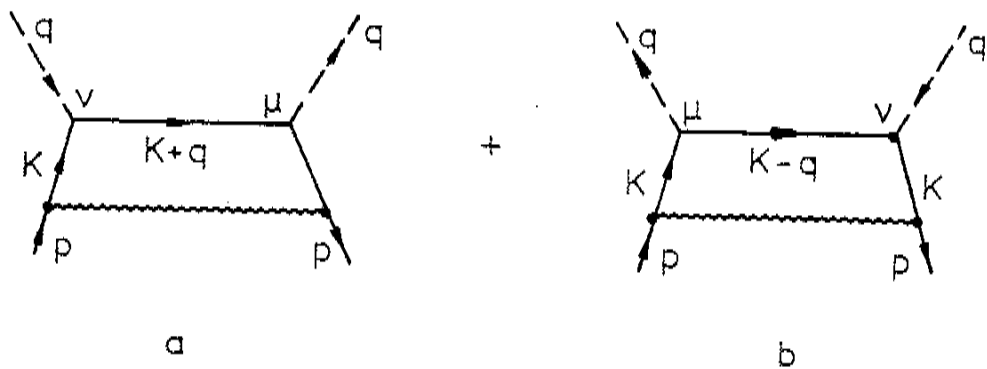
Fig.2: The one-loop approximation for $T_{\mu \nu}$.\\ \\
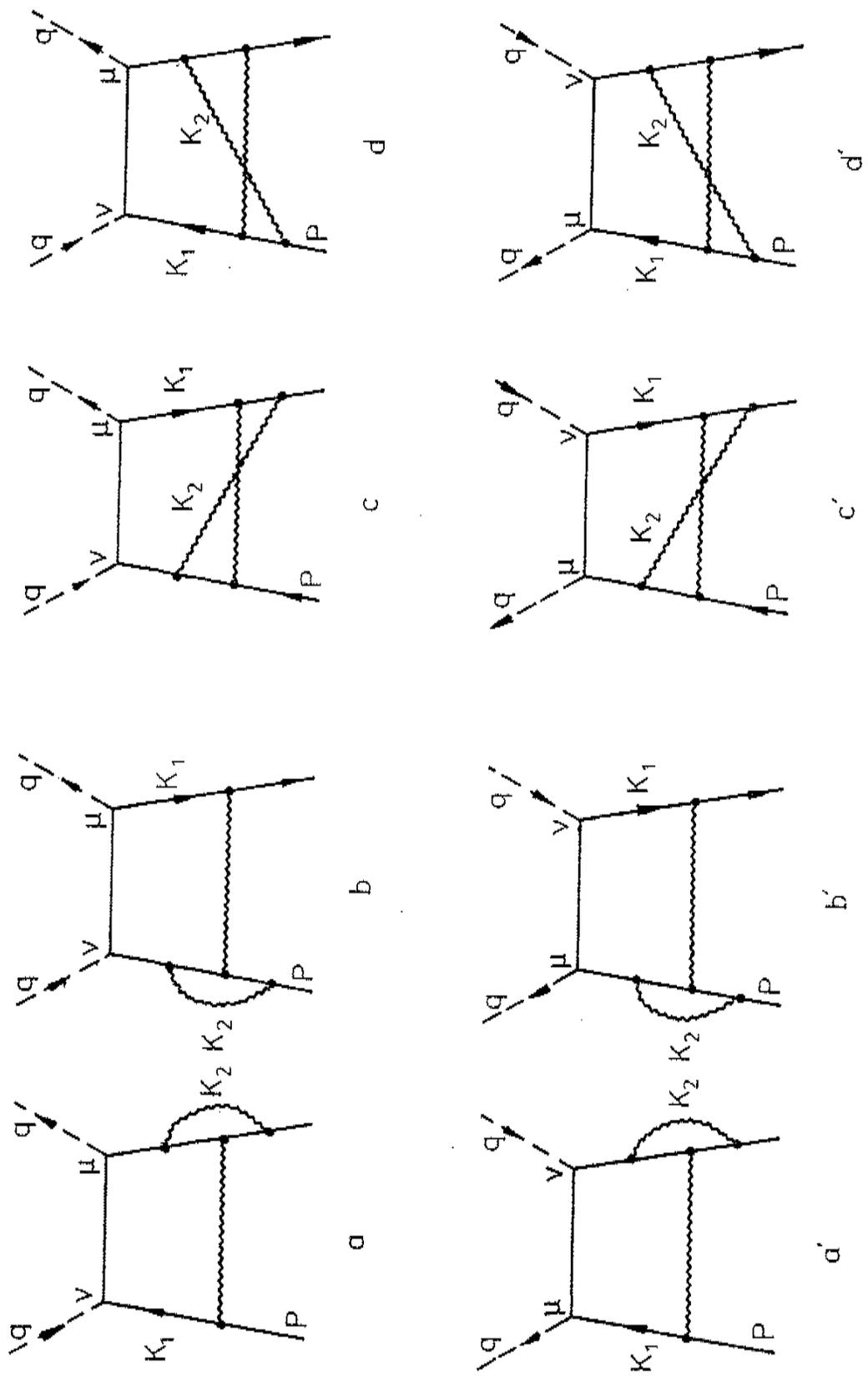
Fig.3: Lowest order non-ladder gluons in $T_{\mu \nu}$.\\ \\
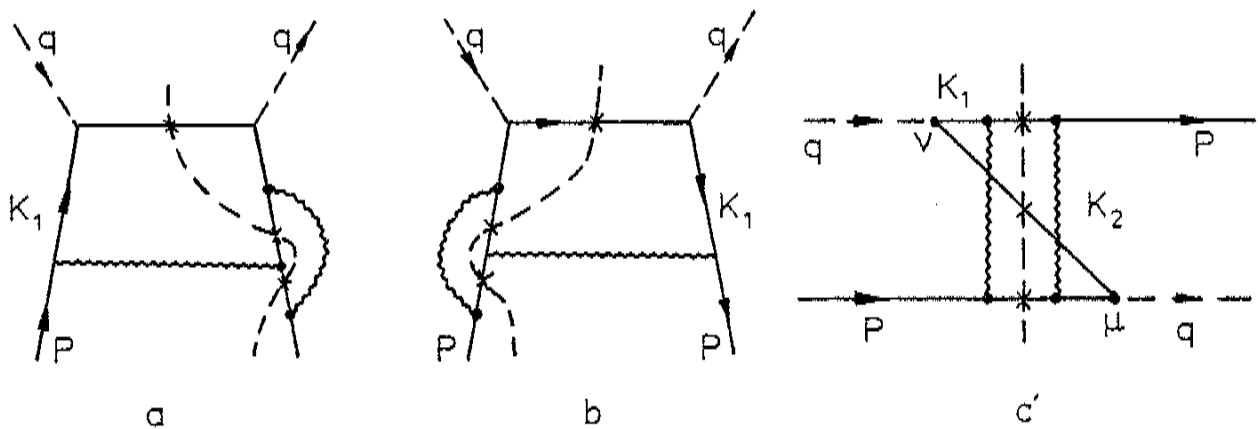
Fig.4: The right hand cut energy discontinuities of the diagrams
Figs.3a, b, and c': in the even signature amplitude, a and c' cancel
(similary a and d', which is not shown in Fig.4), whereas in the odd
signature case they are coming with the same sign and add up.\\ \\
Fig.5: The line with lowest $k_t$ in a general non-ladder diagram for
the quark quark scattering amplitude, (a) a quark line, (b) a gluon
line. (c) the infrared evolution equation for the sum of all non-ladder
diagrams. The sum in front of the second term on the rhs denotes the
different possibilities of attaching bremsstrahlung gluons to the
external lines. \\ \\
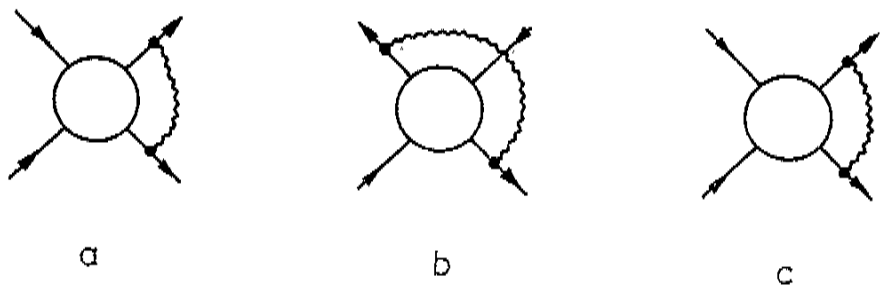
Fig.6: Cancellation of bremstrahlung gluons in the even signature
amplitude.\\   \\
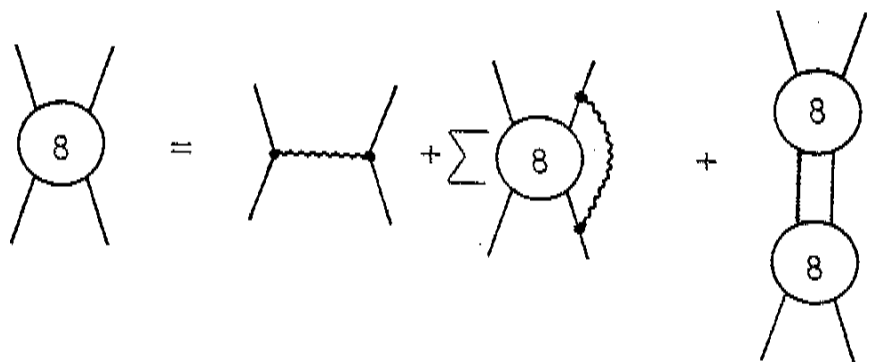
Fig.7: The infrared evolution equation for the octet even signature
quark quark scattering amplitude. The sum in front of the second term
on the rhs denotes the different possibilites ways of attaching
bremsstrahlungs gluons to the external lines. \\ \\
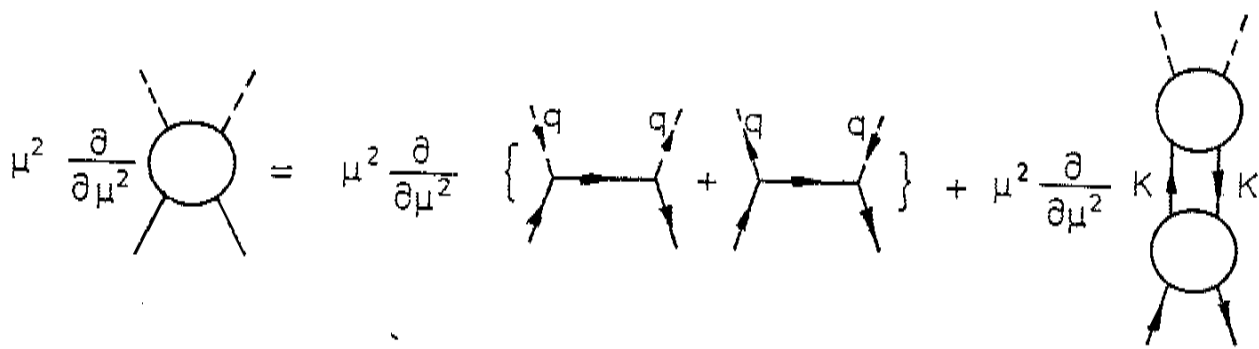
Fig.8: The infrared evolution equation for the photon quark scattering
amplitude. The lower blob on the rhs denotes the quark quark scattering
amplitude from Fig.5c.\\   \\

\end{document}